\documentclass[a4paper]{jpconf}
\usepackage{graphicx}
\usepackage{amsmath,amssymb}
\usepackage{slashed}
\usepackage{color}
\usepackage{subfig}

\newcommand\hc{\text{h.c.}}

\begin{document}
\title{Neutrino spectrum in SU$(3)_\ell\times$SU$(3)_E$ gauged lepton flavor model}

\author{W Sreethawong$^{1\circ}$, W Treesukrat$^{1\ast}$ and P Uttayarat$^{2\dagger}$}

\address{$^1$ School of Physics, Suranaree University of Technology, Nakhon Ratchasima 30000, Thailand.}
\address{$^2$ Department of Physics, Srinakharinwirot University, Wattana, Bangkok 10110, Thailand.}

\ead{$^\circ$warintorn.sut@gmail.com,$^\ast$w.treesukrat@gmail.com,$^\dagger$patipan@g.swu.ac.th}

\begin{abstract}
Massive neutrino is an evidence of new physics beyond the Standard Model. 
One of the well motivated new physics scenarios is a model with gauged lepton flavor symmetry. 
We investigate neutrino properties in the minimal SU$(3)_\ell\times$SU$(3)_E$ gauged lepton flavor model.
In this model, three new species of fermions are introduced to cancel gauge anomalies.
These new fermions lead to a see-saw mechanism for neutrino mass generation. 
We impose the constraints from perturbative unitarity in 2-2 scattering processes, as well as current experimental constraints, to obtain viable neutrino spectrum. We determine the lower bound, with the SU(3)$_\ell$ gauge coupling set to 1, on the lightest neutrino mass of $3.76\times10^{-3}\,(18.9\times10^{-3})\,$ eV for the normal (inverted) hierarchy.\\
\\
This is a proceeding for Siam Physics Congress 2018, May 21-23 2018, Phitsanulok, Thailand.
\end{abstract}

\section{Introduction}
The neutrino has long fascinated physicists. Even though the neutrino was discovered in 1956~\cite{Cowan103}, little is known about its properties. The discovery of neutrino oscillation~\cite{PhysRevLett.87.071301,PhysRevLett.81.1562} established that neutrinos have mass, albeit tiny one. This is in contradiction with the Standard Model (SM) of particle physics. Hence massive neutrino is an evidence that there must be physics beyond SM.  

Additionally, SM contains other theoretical puzzles: Why is the fermion flavor structure the way it is? Why are there three families of quarks and leptons? What is the mechanism behind very light neutrino? A promising framework that could explain these puzzles is the one in which flavor is promoted to gauge symmetry~\cite{Grinstein:2010ve,Alonso:2016onw}. 
Thus, flavor structure is tied to the dynamic that breaks flavor symmetry, and the number of families is related to the size of their representation. 

Since our main interest is the neutrino, in this work we will focus on the gauged lepton flavor symmetry of Ref.~\cite{Alonso:2016onw}. We will take the lepton flavor symmetry to be SU$(3)_\ell\times$SU$(3)_E$ which acts on the electroweak lepton doublets and singlets respectively. The gauging of lepton flavor introduces new gauge anomalies which must be canceled. This requires 3 extra sets of fermions to be introduced. One of the new fermions acts as a right-handed neutrino which then generates a small neutrino mass via a see-saw mechanism~\cite{Mohapatra:1979ia,Schechter:1980gr}. The purpose of this work is to study neutrino properties in the framework of the SU$(3)_\ell\times$SU$(3)_E$ gauged flavor symmetry.

\section{The Model}


\begin{table}[h]
\caption{Transformation properties of SM, extra fermions and flavon fields under the electroweak (the first two rows) and the lepton flavor (bottom two rows) gauge group.}
\centering
\begin{tabular}{@{}*{9}{l}}
\br
& $l_L$ &$e_R$ &$H$ &$\mathcal{E}_R$ &$\mathcal{E}_L$ &$\mathcal{N}_R$ &$\mathcal{Y}_E$ & $\mathcal{Y}_N$ \\
\mr
SU$(2)_L$ &2 & 1 &2 & 1 & 1 &1 &1 &1 \\
U$(1)_Y$ &-1/2 & -1 & 1/2 & -1 & -1 & 0 &0 &0 \\
\mr
SU$(3)_l^{\phantom{i}}$ & 3 & 1 &1 & 3 & 1 &3 &$\bar{3}$ &$\bar6$\\
SU$(3)_E$ & 1 & 3 &1 & 1 & 3 &1 &3 &1\\ 
\br
\label{table:tranf}
\end{tabular}
\end{table}
 
In order to successfully gauge SU$(3)_\ell\times$SU$(3)_E$ flavor symmetry, two scalar fields (flavons), $\mathcal{Y}_E$ and $\mathcal{Y}_N$, and three fermion fields  $\mathcal{E}_R$, $\mathcal{E}_L$ and $\mathcal{N}_R$ are introduced. Their transformation properties are listed in Table~\ref{table:tranf}.  
Their interactions are encoded in the Lagrangian
\begin{equation}
	\mathcal{L} \supset i \bar\psi\slashed{D}\psi + \Tr\left[D_\mu\mathcal{Y}(D^\mu\mathcal{Y})^\dagger\right] + \mathcal{L}_{Yuk} + V(H,\mathcal{Y}_E,\mathcal{Y}_N),
\end{equation}
where $\psi$ are the fermion fields and $\mathcal{Y}$ the flavons. $D$ is the covariant derivative, e.g.
\begin{equation}
	D_\mu\mathcal{N}_R = (\partial_\mu + ig_\ell A^\ell_\mu)\mathcal{N}_R, \qquad
	D_\mu\mathcal{Y}_N = \partial_\mu\mathcal{Y}_N - ig_l(A^\ell_\mu)^T\mathcal{Y}_N -ig_\ell \mathcal{Y}_NA_\mu^\ell,
\end{equation}
where $A_\mu^{\ell(E)}=A_\mu^{\ell(E),a}T^a$ with $T^a$ the SU(3) generator. For definiteness, we take $T^a$ to be half the Gell-Mann matrices.
The Yukawa interactions and fermion mass terms are
\begin{equation}
	\mathcal{L}_{Yuk} = \lambda_E\overline{\ell_L} H \mathcal{E}_R+\mu_E\overline{\mathcal{E}_L}e_R + \lambda_{\mathcal{E}}\overline{\mathcal{E}_L}\mathcal{Y}_E\mathcal{E}_R + \lambda_\nu\overline{{\ell}_L}\tilde{H}\mathcal{N}_R + \frac{\lambda_N}{2}\overline{\mathcal{N}_R^c}\mathcal{Y}_N\mathcal{N}_R +\hc.
\end{equation}
Both electroweak and flavor symmetries are broken spontaneously by  background of the scalars
\begin{equation}
	H\equiv (v+h)/\sqrt{2},\qquad
	\mathcal{Y}_E\equiv \langle\mathcal{Y}_E\rangle + \phi_E/\sqrt{2},\qquad
	\mathcal{Y}_N\equiv \langle\mathcal{Y}_N\rangle + \phi_N/\sqrt{2}.
\end{equation}
The mass matrices for the charged and neutral leptons are
\begin{equation}
	\begin{pmatrix}0 &\lambda_Ev/\sqrt{2}\\ \mu_E & \lambda_{\mathcal{E}}\mathcal{Y}_E\end{pmatrix} +\hc,\qquad
	\frac12\begin{pmatrix}0 &\lambda_\nu v/\sqrt{2}\\ \lambda_\nu v/\sqrt{2} & \lambda_{N}\mathcal{Y}_N\end{pmatrix} +\hc,
\end{equation}
where each entry is a 3 by 3 matrix. Both mass matrices are in a typical see-saw form. Taking $\mathcal{Y}_E\gg v,\mu_E$ and $\mathcal{Y}_N\gg v$, we find the mass matrices for the light leptons, $m_{l(\nu)}$, and the heavy leptons, $M_{\mathcal{E}(N)}$, satisfy
\begin{equation}
	M_\mathcal{E}\simeq \lambda_{\mathcal{E}}\mathcal{Y}_E,\quad
	m_lM_\mathcal{E} \simeq \lambda_{E}\mu_Ev/\sqrt{2},\quad\text{and}\quad
	M_N \simeq \lambda_N\mathcal{Y}_N,\quad
	m_\nu M_N\simeq\lambda_\nu^2v^2/2.
\end{equation}
Working in the basis where $\mathcal{Y}_N$ is diagonal, we deduce
\begin{equation}
	\mathcal{Y}_N \simeq \frac{\lambda_\nu^2v}{2\lambda_N}\text{diag}\left(\frac{v}{m_{\nu_1}},\frac{v}{m_{\nu_2}},\frac{v}{m_{\nu_3}}\right),\quad
	\mathcal{Y}_E \simeq \frac{\lambda_E\mu_E}{\sqrt{2}\lambda_{\mathcal{E}}}U^\dagger\text{diag}\left(\frac{v}{m_e},\frac{v}{m_\mu},\frac{v}{m_\tau}\right)U,
	\label{eq:Y}
\end{equation}
where $U$ is the lepton mixing matrix. Without loss of generality, we take $m_{\nu_1}<m_{\nu_2}<m_{\nu_3}$. Thus $N_1(N_3)$ is the heaviest (lightest) heavy neutrino.

The mass matrix of the flavor gauge bosons is in block diagonal form $\begin{pmatrix}M^2_{\ell\ell} &M^2_{\ell E}\\ M^2_{E\ell} &M^2_{EE}\end{pmatrix}$
where 
\begin{equation}
\begin{split}
	(M^2_{\ell\ell})_{ab} &= g_\ell^2\left[\Tr\left(\mathcal{Y}_E\{T^a,T^b\}\mathcal{Y}_E^\dagger\right) + \Tr\left(\mathcal{Y}_N\{T^a,T^b\}\mathcal{Y}_N^\dagger\right) + \Tr\left(\mathcal{Y}_N^\dagger\{T^{a\,T},T^{b\,T}\}\mathcal{Y}_N\right)\right.\\
	&\qquad\left.+ 2\Tr\left(\mathcal{Y}_N^\dagger T^{a\,T}\mathcal{Y}_NT^b + \mathcal{Y}_N^\dagger T^{b\,T}\mathcal{Y}_NT^a\right)\right],\\
	(M^2_{\ell E})_{ab} &= (M^2_{E\ell})_{ba} = -2g_\ell g_E\Tr\left(T^a\mathcal{Y}_E^\dagger T^b\mathcal{Y}_E\right),\qquad
	(M^2_{EE})_{ab} = g_E^2\Tr\left(\mathcal{Y}_E^\dagger\{T^a,T^b\}\mathcal{Y}_E\right).
\end{split}
\end{equation}
Since $\mathcal{Y}_N\gg \mathcal{Y}_E$, $A^{\ell,a}_\mu$ are approximately the heaviest heavy flavor gauge bosons. 
In the limit $\mathcal{Y}_E\to0$, the $A^{\ell,a}_\mu$ mass matrix gives pairwise eigenvalues (in the unit of $\frac{g_l^2 v^4 \lambda_\nu^4}{4\lambda_N^2 m_{\nu_3}^2}$)
\begin{equation}
\begin{aligned}
	\hat M_{A^1}^2 &= \hat M_{A^2}^2 + 2xy = x^2+xy+y^2,\\ 
	\hat M_{A^4}^2 &= \hat M_{A^5}^2 + 2x = x^2+x+1,
\end{aligned}\quad
\begin{aligned}
	\hat M_{A^6}^2 &= \hat M_{A^7}^2 + 2y = y^2+y+1,\\ 
	\hat M_{A_{\pm}}^2 &=  x^2+y^2+1 \pm {\scriptstyle\sqrt{x^4+y^4 - x^2y^2 -x^2-y^2+1}},
\end{aligned}
\label{eq:gaugebosonmass}
\end{equation}
where $x\equiv m_{\nu_3}/m_{\nu_1}$, $y\equiv m_{\nu_3}/m_{\nu_2}$. $A^{\ell,\pm}_\mu$ are mass eigenstates with  
\begin{equation}
	\begin{pmatrix}A^{\ell,-}_\mu\\A^{\ell,+}_\mu\end{pmatrix} = \begin{pmatrix}c_\alpha &s_\alpha\\ -s_\alpha &c_\alpha\end{pmatrix}\begin{pmatrix}A^{\ell,3}_\mu\\A^{\ell,8}_\mu\end{pmatrix}, \quad
	s_\alpha = \sqrt{\frac12 + \frac{x^2+y^2-2}{4 \sqrt{x^4+y^4-x^2y^2-x^2-y^2+1}}}.
\end{equation}

\section{Partial Wave Unitary Constraints}
\label{sec:unitary}
\begin{figure}
       \centering
        \subfloat{\includegraphics[width= 0.3\textwidth]{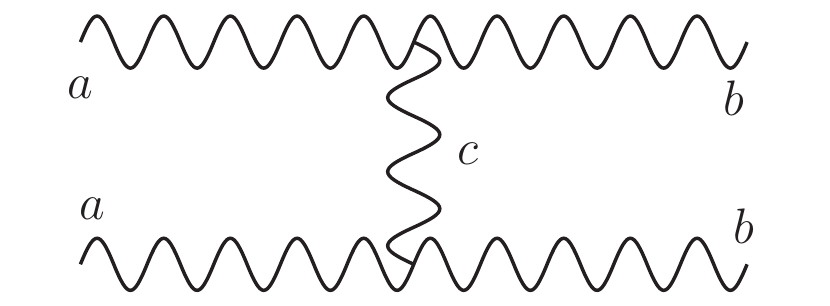}}
        \hspace{.2cm}
        \subfloat{\includegraphics[width= 0.3\textwidth]{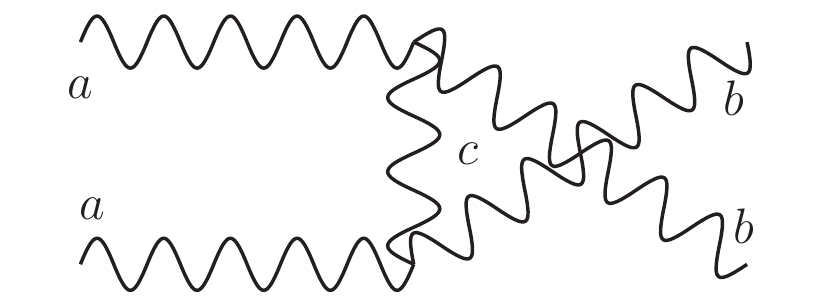}}
        \caption{$t$- and $u$-channel Feynman diagrams contributing to the $A^a_LA^a_L\to A^b_LA^b_L$.}
        \label{fig:2to2}
\end{figure}
The parameter space of the SU$(3)_\ell\times$SU$(3)_E$ gauged lepton flavor model is very large. We will impose perturbative unitary constraint on the 2-2 scattering processes to obtain the viable parameter space. Such a method proved fruitful in constraining the Higgs boson mass~\cite{Lee:1977yc,Lee:1977eg} or the mass of the additional Higgs bosons~\cite{Grinstein:2013fia}. In particular, we will focus on $A^a_LA^a_L\to A^b_LA^b_L$ process where the subscript $_L$ refers to the longitudinal polarization. The relevant Feynman diagrams are shown in Figure~\ref{fig:2to2}. The scattering amplitude is expanded in partial wave as $\mathcal{M} = 16\pi\sum_J(2J+1)a_JP_J(\cos\theta)$  
where $P_J(x)$ is the Legendre polynomial of order $J$ and $\theta$ is the scattering angle. Unitarity of partial wave amplitude requires
$|a_J|\le 1$ and $|\text{Re}(a_J)| \le 1/2$. 
%

Each Feynman diagram contributing to $A^a_LA^a_L\to A^b_LA^b_L$ gives an amplitude that grows with the center-of-mass energy, $\sqrt{s}$. Gauge invariance ensures that the $s^2$ and $s$ growth in the full amplitude cancels. We will focus on the $s^0$ part of the amplitude. In particular we will concentrate on the $J=0$ partial wave amplitude. Taking the $s\to\infty$ limit and subtract off the divergence in the forward direction, we have
\vskip -.4cm
\begin{equation}
\begin{aligned}
a_0 = g^2_\ell &\sum_{c} f^{abc}\left(\scriptstyle\frac{\hat M_{A^a}^4}{\hat M_{A^b}^2 \hat M_{A^c}^2}+\frac{\hat M_{A^b}^4}{\hat M_{A^a}^2 \hat M_{A^c}^2} + \frac{10 \hat M_{A^a}^2}{\hat M_{A^b}^2} + \frac{2 \hat M_{A^b}^2}{\hat M_{A^a}^2}-\frac{\hat M_{A^a}^2}{\hat M_{A^c}^2}-\frac{8 \hat M_{A^a}}{\hat M_{A^b}}+\frac{8 \hat M_{A^b}}{\hat M_{A^a}}-\frac{\hat M_{A^b}^2}{\hat M_{A^c}^2}\right),
	\label{eq:unitary}
\end{aligned}
\end{equation}
where only the index $c$ is summed over, $f^{abc}$ is the SU(3) structure constant, $\hat M^2_{A^a}$ is given in Equation~\eqref{eq:gaugebosonmass} and we have dropped terms suppressed by $\mathcal{O}(1/s)$. For the case $a=\pm$, we have $T^+ = s_\alpha T^3+c_\alpha T^8$ and $T^- = c_\alpha T^3-s_\alpha T^8$.
Since the partial wave amplitude is real for any $a,b$, we impose the constrain Re$[a_0]\le1/2$.


\section{Results}
Neutrino oscillations place constraints on the neutrino squared mass difference. In the normal hierarchy scenario, $m_{\nu_1}\lesssim m_{\nu_2}\ll m_{\nu_3}$, the constraints are $m_{\nu_3}^2-m_{\nu_1}^2\in [2.45,2.69]\times10^{-3}\text{eV}^2$ and $m_{\nu_2}^2-m_{\nu_1}^2\in [6.93,7.96]\times10^{-5}\text{eV}^2$ at 99\% confidence level (CL)~\cite{Olive:2016xmw}. For the case of inverted hierarchy, $m_{\nu_1}\ll m_{\nu_2}\lesssim m_{\nu_3}$, the constraints are $m_{\nu_3}^2-m_{\nu_1}^2\in [2.42,2.66]\times10^{-3}\text{eV}^2$ and $m_{\nu_3}^2-m_{\nu_2}^2\in [6.93,7.96]\times10^{-5}\text{eV}^2$ at 99\% CL~\cite{Olive:2016xmw}. Moreover, the sum of the three neutrino masses is constrained by cosmological observations. Combining constraints from the cosmic microwave background and baryonic acoustic oscillation gives $\sum m_{\nu_i}\le0.17$ eV at 95\% CL~\cite{Olive:2016xmw}.

\begin{figure}[h]
\includegraphics[width=19pc]{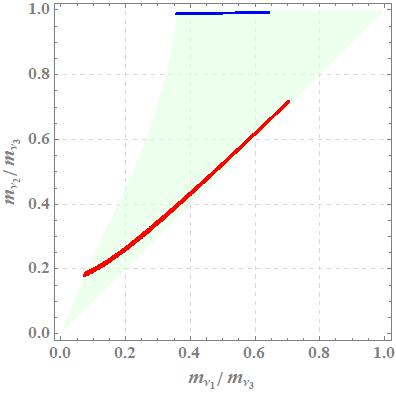}\hspace{2pc}
\begin{minipage}[b]{16pc}
\caption{Viable neutrino spectrum for normal hierarchy (red) and inverted hierarchy (blue). Region compatible with perturbative unitarity is shown in green.
This translates to the lower bound on the lightest neutrino mass $m_{\nu_1}\ge3.76\times10^{-3}\,(18.9\times10^{-3})\,$eV in the normal (inverted)}
\label{fig:spectrum}
\end{minipage}
\end{figure}


Combining the above experimental constraints with the constraint from partial wave unitary, Equation~\eqref{eq:unitary}, we obtain the viable neutrino spectrum for the case $g_\ell=1$, see Figure~\ref{fig:spectrum}.
\section{Conclusions and Final Remarks}
Experimental constraints on neutrino mass alone cannot determine the lower bound on the lightest neutrino mass, $m_{\nu_1}$. Combining experimental and partial wave unitary constraint, we determine the lower bound on $m_{\nu_1}$. In the context of the gauged SU$(3)_\ell\times$SU$(3)_E$ lepton flavor with the gauge coupling $g_\ell=1$, we find $ m_{\nu_1}\ge3.76\times10^{-3}\,(18.9\times10^{-3})\,$ eV in the normal (inverted) hierarchy scenario.

In principle, one could also obtain unitary constraints from the $A^a_LA^a_L\to N_i\bar N_i$ and $N_i\bar N_i\to N_j\bar N_j$ processes. However, we have checked explicitly that these constraints are automatically satisfied.
\ack
The work of WS has been supported in part by Suranaree University of Technology under grant no. 72/2559 SUT1-105-59-12-12.
The work of PU has been supported in part by the Thailand Research Fund under contract no.~ MRG6080290, the National Science and Technology Development Agency of Thailand under grant SCHNR-2015-841, and the Faculty of Science, Srinakharinwirot University under grant no.~422/2560. 
\section*{References}
\bibliography{reference}\bibliographystyle{iopart-num}

\end{document}